# Induced Current Due to Electromagnetic Shock Produced by Charge Impact on a Conducting Surface

D. Li[1], P. Wong[2], D. Chernin[3], and Y. Y. Lau[1]

[1]*Department of Nuclear Engineering and Radiological Sciences, University of Michigan, Ann Arbor, Michigan 48109, USA*
[2]*Department of Electrical and Computer Engineering, Michigan State University, East Lansing, Michigan 48824, USA*
[3]*Leidos Inc., Reston, Virginia 20190, USA*

**Abstract**: This paper compares the transient induced current due to the electromagnetic shock produced by a charged particle impacting a perfectly conducting plate, with the classical, quasi-static induced current of Ramo and Shockley (RS). We consider the simple model of a line charge, removed upon striking the plate. We find that the induced current due to the shock is negligible compared with the RS current for nonrelativistic impact energies, but is more significant as the impact energy becomes mildly relativistic. The implications of these findings are discussed.

## I.      Introduction

The classical Ramo-Shockley (RS) theorem gives the current induced on perfect conductors by the motion of nearby charges, assuming nonrelativistic motion of those charges in electrostatic fields [1], [2]. The RS theorem has been used with great success in vacuum electronics to approximate the radio frequency (RF) beam current that drives traveling wave tubes, klystrons, and crossed-field devices [3], [4], [5], [6]. It has also been used in the theory of multipactor discharge, to analyze its effects on the beam loading of RF circuits [7], and on the degradation in signal quality [8]. We recently suggested the reasons for such successes, having extended the RS theory to include electromagnetic and relativistic effects, for the first time [9]. We obtained an exact, closed form analytic solution for the induced current of a simple model of a line charge moving between two parallel plates, and compared that solution with the classical RS value. The classical RS approach does not account for electromagnetic transients and multiple reflections, but otherwise works quite well even at relativistic velocities, before the electron line charge strikes and is removed by a conducting surface, at which time an electromagnetic shock is generated. This shock, a form of transition radiation [10], is completely absent from the electrostatic theory of RS. Note that the classical RS theory gives a zero value of the induced current at the instant of impact, and thereafter. Thus, the additional beam loading due to the impact of electrons was omitted in all previous analyses of beam loading that made use of the classical RS. This paper addresses this issue by comparing the magnitude of the induced current produced by the transient electromagnetic shock due to the disappearance of a charge upon impact on a perfectly conducting plate, with the classical induced current according to RS.

In this paper, we will continue to use the simple model of a line charge [9] moving between two parallel conducting plates of separation $d$ (Fig. 1a). We will further assume that an AC voltage $V_{RF}$ of frequency $\omega$ is imposed across the parallel plates and we use the classical RS theory to calculate the induced current due to the motion of the line charge within the plates during its transit, i.e., before the line charge strikes a plate. We assume that the line charge motion is subject only to the vacuum AC electric field, ignoring the additional electrostatic field due to the image charges [7], [11]. We consider only the motion of the line charge in one transit, from one plate to the opposite plate, with a transit time $T = \pi/\omega$ exactly equal to 1/2 of the RF period. This assumption is consistent with the sinusoidal steady state assumption, where an identical trip is repeated every cycle. The classical induced current according to RS during this single transit may then be calculated. When the line charge hits a plate, it is removed, and the classical RS induced current is identically zero. To assess the electromagnetic shock-induced current due to the impact of the line charge at the terminal velocity $v_z$, we consider a separate, reduced problem in which we assume a single plate geometry, $d \to \infty$ (Fig. 1b). We assume that the line charge starts at a large distance from this single plate and it moves toward this plate at the constant terminal velocity, $v_z$, which is set equal to the impact velocity in the parallel plate model. This line charge hits the single plate and is instantaneously removed. The shock-induced solution in this single plate problem may then be extracted from our exact solutions previously obtained. We find that this shock induced current, when compared with the classical RS induced current during transit (see Fig. 4 below), is relatively unimportant when $\beta = v_z/c$ is small, but becomes significant as $\beta$ increases. These findings suggest that the electromagnetic shock would not significantly worsen the signal quality in a multipactor discharge, an important issue for satellite communication, but it could affect the beam loading in a relativistic magnetron or a magnetically insulated line oscillator (MILO). We shall further discuss these issues in Section III.

Section II presents the calculation of the electromagnetic, shock-induced current in a single-plate geometry, assuming that a line charge approaches this plate at a constant speed $v_z$ from a faraway distance, using the model in Fig. 1b. A calculation of the electrostatic induced current during the transit of a line charge that moves between two parallel plates that is subjected to the sinusoidal gap voltage, $V_{RF}$, using the model in Fig. 1a, is given in Section III. This electrostatic induced current is compared with the shock-induced current calculated in Section II, assuming an identical impact velocity, $v_z$. We also discuss the implications from this comparison. Section IV provides a summary and some additional observations.

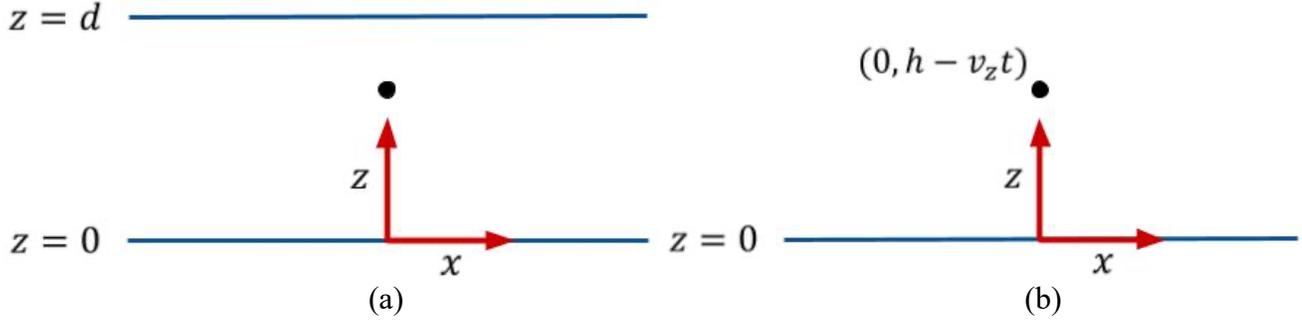

**Figure 1.** (a) Parallel plate model used to assess the electrostatic induced current. A sinusoidal steady state voltage is imposed across the plates. (b) Single-plate model used to assess the shock-induced current. The line charge lies at $(x, z) = (0, h)$ initially. At $t = 0$, the charge experiences an impulse, downward acceleration to a velocity $v_z$. We shall later take $h \to \infty$, $T = h/v_z \to \infty$ and $t \to \infty$, with $(t - T)$ finite, to isolate the shock-induced current.

## II.     Shock-Induced Current in the Single Plate Geometry

In this section, we evaluate the shock induced current when a line charge of charge density ($\lambda$, in C/m) strikes a perfectly conducting plate with velocity $v_z$, and is removed from the surface on impact. To avoid the complexities of multiple reflections of the electromagnetic waves, we consider the single plate geometry of Fig. 1b. We assume that the line charge is initially at rest and is located at (x, z) = (0, h), as shown in Fig. 1b. At t = 0, an impulse acceleration of the line charge is applied, so that the line charge moves downward at a constant velocity $-v_z$, striking the single plate at time $t = T = h/v_z$. For $0 < t < T$, the induced current on the conducting plate is due to the impulse acceleration at t = 0, and to the re-arrangements of the surface charge as a result of the constant downward motion of the line charge. The exact expression for the induced current on the conducting plate, $K_x = K_x(x, t)$, is given by Eq. (8) of [9],

$$K_x = \frac{\lambda \gamma^2 v_z x}{\pi[x^2 + \gamma^2(h - v_z t)^2]} \times \frac{ct - hv_z/c}{[(ct)^2 - x^2 - h^2]^{1/2}}, \quad t < T, \tag{1}$$

where we have replaced $v_z \to -v_z$ with $v_z > 0$. In Eq. (1), $\gamma = (1 - \beta^2)^{-1/2}$, $\beta = v_z/c$ and $c$ is the speed of light.

To extract the shock-induced solutions, we need to eliminate the transient component, and to pay special attention to the solution in the vicinity of $t = T$. This may be accomplished by defining $t' = t - T$ and taking the limit of $t, T \to \infty$, while $t'$ is finite. Physically, we assume $h \to \infty$ so that when the line charge hits the plate, the transient solution due to the impulse acceleration disappears. We then obtain (see Appendix A)

$$\bar{K}_x = \frac{\gamma \bar{x}}{\pi[\bar{x}^2 + \gamma^2 \bar{t}'^2]} \tag{2}$$

for $t' < 0$. We have defined the dimensionless quantities $\bar{x} = x/L$, $\bar{t} = v_z t/L$, and $\bar{K} = K/(\lambda v_z/L)$ where $L$ is an arbitrary scale length.

We next consider the induced current for all time, including $t \geq T$, after the line charge strikes the conducting plate and is removed instantaneously. In this case, the induced current may be derived from Eq. (14) of [9], which may be written, upon taking the single plate limit ($n = 0$) and, once again, taking $\bar{h}, \bar{t} \to \infty$, while $\bar{t}'$ is finite, (see Appendix B)

$$\bar{K}_x = \begin{cases} \dfrac{\gamma \bar{x}}{\pi[\bar{x}^2 + \gamma^2 \bar{t}'^2]}, & \bar{t}' < 0 \\ \dfrac{\gamma \bar{x}}{\pi[\bar{x}^2 + \gamma^2 \bar{t}'^2]} \left\{ 1 - \dfrac{\gamma \bar{t}'/\beta}{[(\bar{t}'/\beta)^2 - \bar{x}^2]^{1/2}} \right\}, & \bar{t}' \geq 0 \end{cases}$$

(3a,b)

where we note that Eq. (3a) is identical to Eq. (2). It is straightforward to show that in the quasi-static limit, or when $c \to \infty$ or $\gamma \to 1$, Eq. (3) reduces to the classical RS result,

$$\bar{K}_x^{ES} = \begin{cases} \dfrac{\bar{x}}{\pi[\bar{x}^2 + \bar{t}'^2]}, & \bar{t}' < 0 \\ 0, & \bar{t}' \geq 0 \end{cases}$$

(4)

where the superscript *ES* stands for "electrostatic".

Since we are only interested in the induced current due to the electromagnetic shock produced by the line charge striking the conducting plate, we may modify Eq. (3) to be nonzero strictly within the "shock cone", $|x| < c(t - T)$, of the radiation produced by the impacting line charge. This allows us to write the induced current due exclusively to the shock, ignoring all transients (and all electromagnetic wave reflections for the case of a parallel plate geometry),

$$\bar{K}_x = \bar{K}_x^{SH} = \begin{cases} \dfrac{\gamma \bar{x}}{\pi[\bar{x}^2 + \gamma^2 \bar{t}'^2]} \left\{ 1 - \dfrac{\gamma \bar{t}'/\beta}{[(\bar{t}'/\beta)^2 - \bar{x}^2]^{1/2}} \right\}, & \bar{t}' > |\beta \bar{x}| \\ 0, & \text{otherwise} \end{cases}$$

(5)

where the superscript *SH* stands for "shock-induced". Clearly, $\bar{K}_x^{ES} = 0$ when $\bar{t}' > |\beta \bar{x}|$ according to Eq. (4).

As a sense of scale for the magnitude of the quasistatic induced current, we compare the maximum value of $\bar{K}_x^{ES}$ at $\bar{x} = 1$ with $\bar{K}_x^{SH}$, noting that $\bar{K}_x^{ES}(\bar{x} = 1)$ is largest when $\bar{t}' = 0$ [c.f. Eq. (4)], taking on the value $\bar{K}_x^{ES} = 1/\pi$. The exact induced current at multiple values of $\beta$ are plotted with $-1$ times the reference value $1/\pi$ in Fig. 2. Figure 3 illustrates how the impact velocity of the line charge affects the time-evolution of the shock-induced current. As with Fig. 2, the reference value of $-1/\pi$ is overlaid.

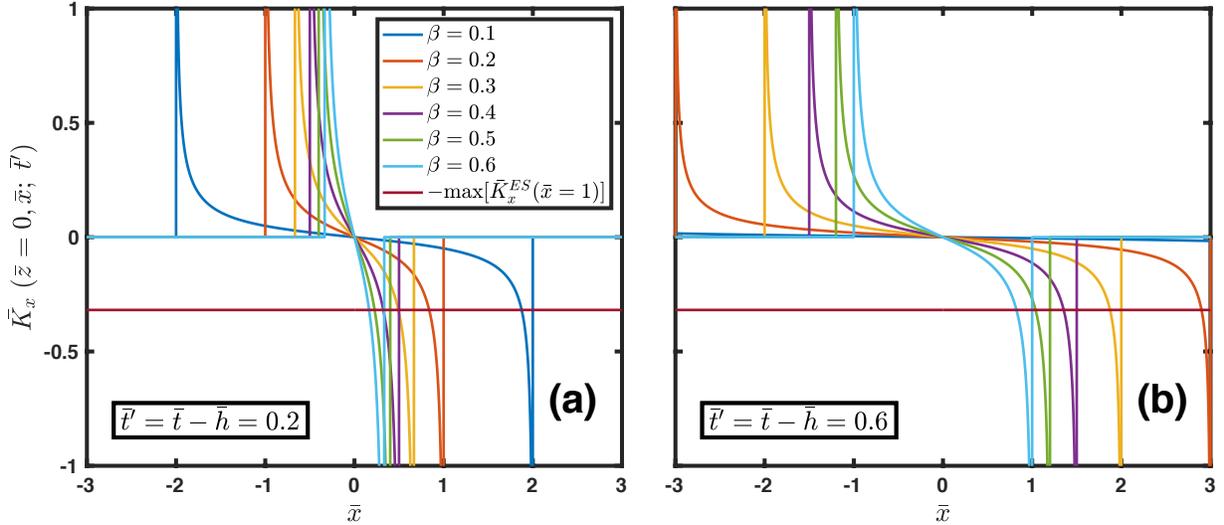

**Figure 2.** The shock-induced currents, $\bar{K}_x^{SH}$, [see Eq. (5)] at $\bar{x} = 1$ evaluated at (a) $\bar{t}' = \bar{t} - \bar{h} = 0.2$ and (b) $\bar{t}' = 0.6$ with six different $\beta$ values. Also plotted for reference is $-1$ times the maximum value of the classical Ramo-Shockley induced current, $\bar{K}_x^{ES}(\bar{x} = 1)$.

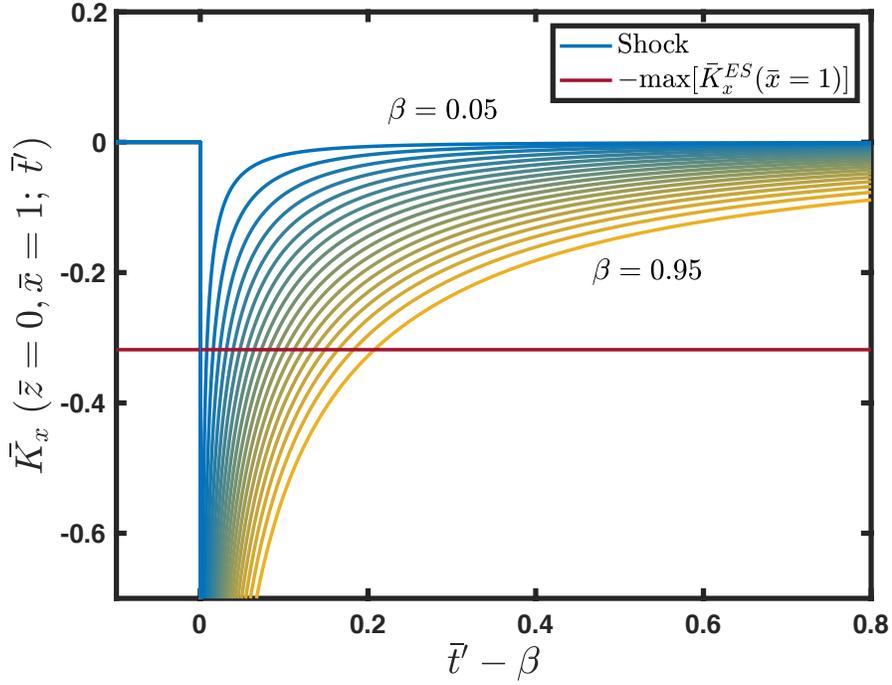

**Figure 3.** Time-evolution of the shock-induced currents, $\bar{K}_x^{SH}$, [see Eq. (5)], evaluated at $\bar{x} = 1$ on the single plate at $z = 0$, due to an impacting line charge for 19 different values of $\beta$ ($\beta = 0.05, 0.1, ..., 0.95$), compared with the reference electrostatic induced current value $1/\pi$. Note that the blue curves correspond to lower (less relativistic) impact velocities and the yellow curves correspond to higher (more relativistic) impact velocities.

It appears that the "widths" of the normalized shock-induced current curves, $\bar{K}_x^{SH}$ vs. $\bar{t}'$ (Fig. 3), are narrower for less relativistic velocities and wider for more relativistic velocities. In the limit $\beta \to 0$, the shock-induced current curve becomes infinitely narrow. Regardless of $\beta$, the shock-induced surface current approaches $\bar{K}_x^{SH} \to -\infty$ at $\bar{t}' - \beta = 0$. However, it appears that for deeply nonrelativistic impact velocities, corresponding to 100's eV or less such as those associated with the participating electrons in a multipactor discharge [7], the shock-induced current would have a negligible influence, compared with the induced current during transit that is described by the classical RS. This comparison is given in the next section.

### III.    Comparison with the Classical Induced Current in a Parallel Plate Geometry

For the induced current according to the classical RS theory, we use the parallel plate model (Fig. 1a) and assume that the potential at the lower plate is zero and at the upper plate is $V_{RF} = \tilde{V}_{RF}\sin(\omega t_1 + \theta)$, where the electron line charge is initialized at $t_1 = 0$ somewhere within the plates. We shall consider only the classical electrostatic induced current during transit, i.e., before this electron strikes a plate. If the initial position of the line charge is at the upper plate, and if it reaches the lower plate after half an RF cycle (i.e., the transit time $\tau = \pi/\omega$), this is equivalent to the model of a 2-surface steady state multipactor discharge of order one [7], [11]. There is a "phase-locking condition" on the sinusoidal steady state. Assuming nonrelativistic voltage and ignoring space charge effects, this condition reads [11]

$$\tilde{V}_{RF} = \frac{md^2}{e} \frac{\omega^2 \left(1 - \frac{\pi v_0}{\omega d}\right)}{2\sin\theta + \pi\cos\theta}$$

(6)

where $\tilde{V}_{RF}$ is the amplitude of the RF voltage, $\theta$ is the "launch phase" and $v_0$ is the initial velocity of the line charge. Setting $v_0 = 0$ and $\theta = \pi$, $\tilde{V}_{RF}$ has a magnitude,

$$\tilde{V}_{RF} = \frac{md^2\omega^2}{\pi e}.$$

(7)

We next solve for $z(t_1)$ and $\dot{z}(t_1)$ from the non-relativistic force law within the parallel plates (Fig. 1a) to obtain

$$z(t_1) = d - \frac{d}{\pi}(\omega t_1 - \sin \omega t_1)$$

(8a)

$$\dot{z}(t_1) = -\frac{\omega d}{\pi}(1 - \cos \omega t_1),$$

(8b)

so that $z(0) = d$ and $\dot{z}(0) = 0$. In the quasistatic regime, the classical (RS) induced current on the lower plate is [9],

$$\bar{K}_x^{ES,RF} = -\frac{\lambda \dot{z}(t_1)}{2d} \frac{\sinh\left(\frac{x\pi}{d}\right)}{\cosh\left(\frac{x\pi}{d}\right) - \cos\left(\frac{\pi}{d}z(t_1)\right)} = \frac{\lambda \omega}{2\pi} \frac{(1 - \cos \omega t_1)\sinh\left(\frac{x\pi}{d}\right)}{\cosh\left(\frac{x\pi}{d}\right) - \cos\left(\frac{\pi}{d}\left(d - \frac{d}{\pi}(\omega t_1 - \sin \omega t_1)\right)\right)}$$

(9)

where $0 < t_1 < \tau$ and $\tau = \pi/\omega$ is the transit time of the line charge. Since we have previously set the magnitude of the impact velocity as the velocity scale, $v_z$, let us do the same here. We thus obtain $\dot{z}(t_1 = \pi/\omega) = -2\omega d/\pi = -v_z$, and similarly non-dimensionalize $\bar{t}_1 = 2\omega t_1/\pi$ and $\bar{K} = K/(2\lambda\omega/\pi)$. The normalized classical induced current becomes,

$$\bar{K}_x^{ES,RF} = \begin{cases} \frac{1}{4}\frac{\left(1 - \cos\left(\frac{\pi}{2}(\bar{t}' + 2)\right)\right)\sinh(\pi\bar{x})}{\cosh(\pi\bar{x}) - \cos\left(-\frac{\pi}{2}\bar{t}' + \sin\left(\frac{\pi}{2}(\bar{t}' + 2)\right)\right)}, & -2 < \bar{t}' < 0 \\ 0, & \text{otherwise} \end{cases}$$

(10)

where $\bar{t}' = \bar{t}_1 - \bar{\tau} = \bar{t}_1 - 2$. Note that Eq. (10) is independent of $\tilde{V}_{RF}$ and $\omega$. This enables convenient comparison between this classical induced current during transit and the shock-induced current. The shock-induced current depends on $\tilde{V}_{RF}$, which determines the impact velocity, $v_z = \beta c$, the velocity assumed in the model in Section II. Figure 4 overlays Eq. (10) with Eq. (5), taking on multiple $\beta$ values. Each $\beta$ value corresponds to a unique $\omega$ and, therefore, a unique $\tilde{V}_{RF}$ by $\omega = \pi c \beta/2d$ and $\tilde{V}_{RF} = \pi m c^2 \beta^2/4e$, respectively, in the nonrelativistic case.

The analysis in this section has thus far assumed non-relativistic electron motion. An extension to relativistic motion using the same sinusoidal AC gap voltage is given in Appendix C. Comparisons of the nonrelativistic and relativistic values of $\omega$ and $\beta$ as a function of $\tilde{V}_{RF}$ are shown in Fig. 5.

It is important to note the comparison in Fig. 4 has ignored the multiple reflections which necessarily occur in the parallel plate model in a fully electromagnetic theory. A fully electromagnetic solution has not been obtained for the AC case. Such a solution is expected to be exceedingly complicated and would easily mask the effects of the shock which is the subject of this paper. Figure 4, on the other hand, gives a much clearer illustration of the order of magnitude of the shock-induced current, compared with the classical induced current according to RS. This comparison was made ignoring all space charge effects.

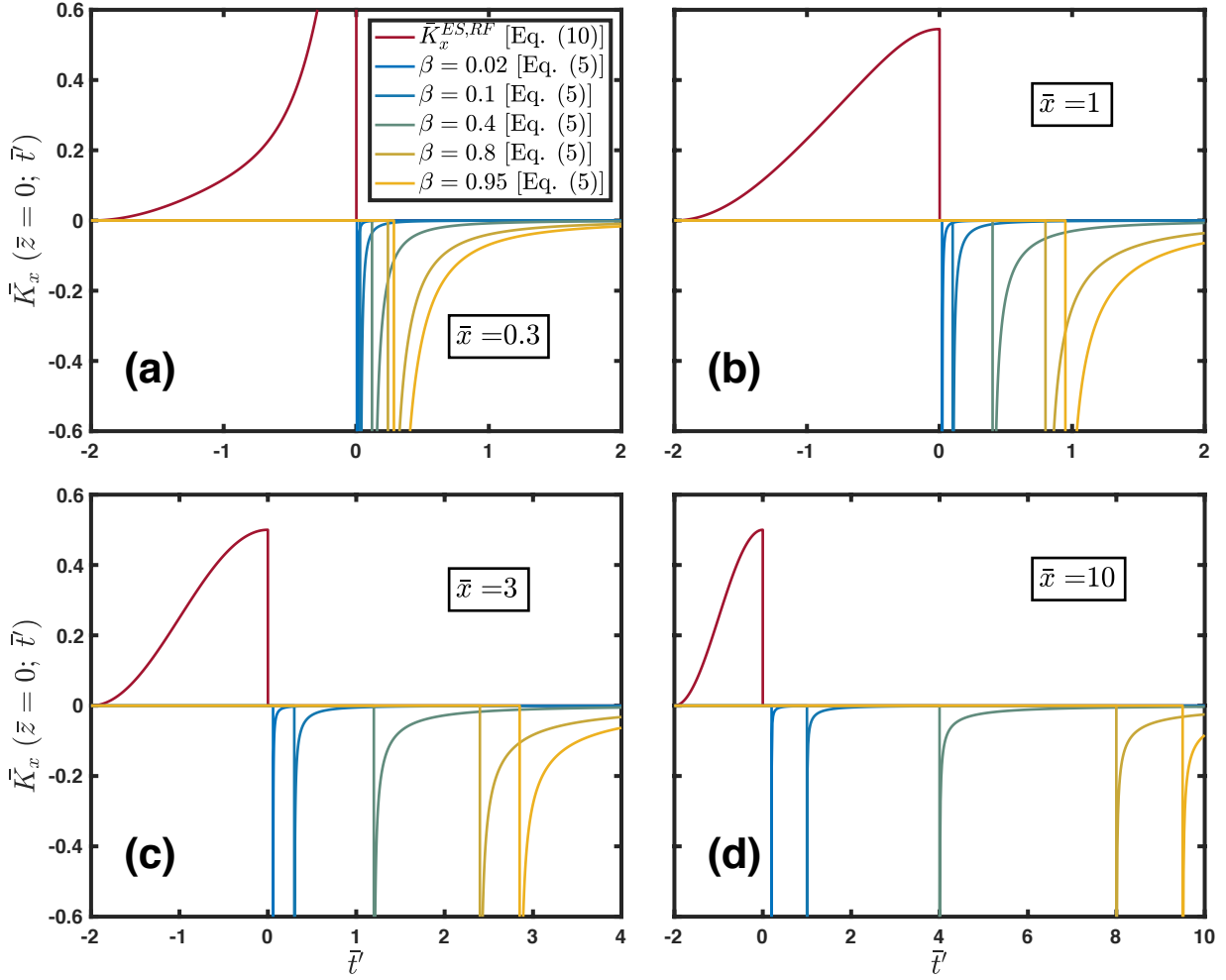

**Figure 4.** Time-evolution of the normalized electrostatic, parallel-plate induced current from a line charge subject to an RF electric field [see Eq. (10)], compared with the normalized induced currents due to the electromagnetic shock of the impacting line charge [see Eq. (5)] at $\bar{x}$ = (a) 0.3, (b) 1.0, (c) 3.0, and (d) 10. Five different values of $\beta$ were used, $\beta = 0.02, 0.1, 0.4, 0.8, 0.95$, corresponding to impact energies of, respectively, 0.102, 2.574, 46.54, 340.7, and 1126 keV. The normalized electrostatic induced current exists only for $\bar{t}' < 0$ and is independent of the gap voltage (or $\beta$) in the present model. The shock-induced current exists only for $\bar{t}' \geq 0$.

Figure 4 allows us to draw the following inferences on the roles of the shock-induced current, in comparison with the classical induced current according to RS.

(i) The shock-induced current is unlikely to further degrade the signal quality in satellite communication as a result of multipactor discharge. The underlying reason is that the multipactoring electrons typically have impact energies of order 100 eV or less [7], at which the shock-induced current is too small to add significantly to the in-phase and quadrature components of the signal errors [8].

(ii) The shock-induced current becomes appreciable for high impact energies, e.g., at 340 keV ($\beta = 0.8$) as shown in Fig. 4. Thus, the shock-induced current may provide appreciable additional beam loading to the RF circuits in those radiation sources in which the electrons strike the RF circuit with mildly relativistic energy. A magnetically insulated line oscillator (MILO) and a relativistic magnetron may therefore suffer from appreciable beam loading due to this shock-induced current as the spoke electrons always strike the anode slow wave structure [12], [13].

(iii) Linear beam tubes, such as the traveling wave tubes and klystrons, would not be affected as much from this shock-induced current because there is little beam interception in the interaction region.

(iv) One may wonder if the additional beam loading due to the electromagnetic shocks, as described in (ii) for relativistic crossed-field devices, is one reason for the characteristically low efficiencies experienced in relativistic magnetrons (compared with microwave oven magnetrons) and in MILO experiments [14], [15].

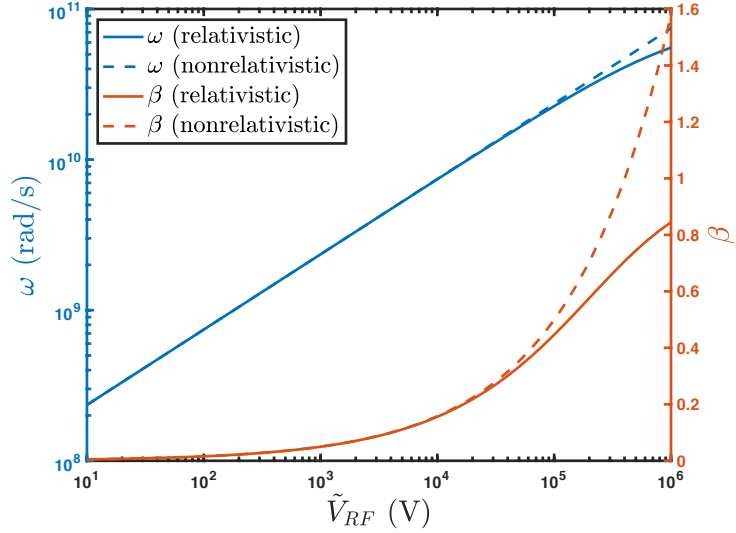

**Figure 5.** Comparison of $\omega$ and $\beta$ as a function of $\tilde{V}_{RF}$ in the nonrelativistic (dashed curves) and relativistic (solid curves) regimes with gap spacing $d = 0.01$ m.

## IV. Concluding Remarks

In this paper, we have isolated the effects of the electromagnetic shock phenomenon produced by the impact of a line charge on a conducting plate and evaluated the associated induced surface current. We have compared the induced current of this purely electromagnetic effect with the electrostatic induced current of a charge subject to an RF voltage in a parallel-plate geometry. This comparison was made only during one transit, which presumably gives an adequate assessment for steady state operation in which one transit is essentially repeated each RF cycle. We have ignored all electromagnetic wave reflections, which necessarily occur in a parallel plate geometry. This also is presumably adequate since the classical RS theory, which ignores all such reflections, is known to give a reasonably accurate value of the induced current in RF vacuum electronics, when there is little beam interception. The underlying reason for this unexpected feature has been explored [9].

We tentatively conclude that the shock-induced surface current will not cause additional damage in multipactor discharge, because of the low impact energies of the multipactoring electrons. However, the shock-induced currents could be important in MILOs and relativistic magnetrons, in which spoke electrons, a significant fraction of which may have high energies (a few 10's to 100's keV), always strike the anode circuit [14], [15]. The shock induced current in this case would introduce additional beam loading which is omitted in the classical RS theory. This additional beam loading, leading to de-tuning and de-Q'ing of the circuit, is not easy to quantify. Here, we only note that they could be appreciable in MILOs and in relativistic magnetrons. We suggest that this additional beam loading is unimportant for non-relativistic crossed-field devices, consistent with our experience that the classical RS induced current provides reasonably accurate results in computer simulations, when compared with measurements [5], [6].

### Acknowledgment

The authors would like to acknowledge useful discussions with Professor Zhong He at the University of Michigan. This work was supported by AFOSR Grant Nos. FA9550-18-1-0153, FA9550-18-1-0062, and FA9550-21-1-0367, and in part by L3Harris Electron Devices Division.

## Appendix A: Derivation of Eq. (2)

With $vt' \equiv vt - h$, we write Eq. (1) as,

$$K_x = \frac{\lambda \gamma^2 vx}{\pi[x^2 + \gamma^2 v^2 t'^2]} \times \frac{ct + \frac{v^2 t' - v^2 t}{c}}{[(ct)^2 - x^2 - v^2 t'^2 - v^2 t^2 + 2v^2 tt']^{1/2}}.$$

(A-1)

We now take the limit $t \to \infty$ (and $t'$ is finite) to get

$$\lim_{t \to \infty} K_x = \frac{\lambda \gamma^2 vx}{\pi[x^2 + \gamma^2 v^2 t'^2]} \times \frac{c - \frac{v^2}{c}}{[c^2 - v^2]^{1/2}} = \frac{\lambda \gamma vx}{\pi[x^2 + \gamma^2 v^2 t'^2]}.$$

(A-2)

In terms of dimensionless quantities defined in the paragraph below Eq. (2), we may write

$$\bar{K}_x = \frac{\gamma \bar{x}}{\pi[\bar{x}^2 + \gamma^2 \bar{t}'^2]}$$

(A-3)

for $\bar{t}' < 0$.

## Appendix B: Derivation of Eq. (3)

Imposing the single plate limit ($n = 0$) and $v \to -v$, we may write Eq. (14b) of [9] as

$$\bar{K}_x = \mathbf{K}(\bar{x}, -\bar{t}, \bar{h}) - \mathbf{K}(\bar{x}, -(\bar{t} - \bar{T}), 0)$$
$$= -\frac{\bar{x}(-\bar{t}/\beta + \bar{t}\beta - \bar{t}'\beta)}{\pi[(\bar{t}/\beta)^2 - \bar{x}^2 - \bar{t}'^2 - \bar{t}^2 + 2\bar{t}\bar{t}']^{1/2}[\bar{t}'^2 + \bar{x}^2/\gamma^2]} + \frac{\bar{x}(-\bar{t}'/\beta)}{\pi[(\bar{t}'/\beta)^2 - \bar{x}^2]^{1/2}[\bar{t}'^2 + \bar{x}^2/\gamma^2]}$$

(B-1)

where $t' = t - T$. Taking the limit of $t, T \to \infty$, while $t'$ is finite, we obtain

$$\lim_{t \to \infty} \bar{K}_x = -\frac{\bar{x}(-1/\beta + \beta)}{\pi[1/\beta^2 - 1]^{1/2}[\bar{t}'^2 + \bar{x}^2/\gamma^2]} + \frac{\bar{x}(-\bar{t}'/\beta)}{\pi[(\bar{t}'/\beta)^2 - \bar{x}^2]^{1/2}[\bar{t}'^2 + \bar{x}^2/\gamma^2]}$$
$$= \frac{\gamma \bar{x}}{\pi[\bar{x}^2 + \gamma^2 \bar{t}'^2]}\left\{1 - \frac{\gamma \bar{t}'/\beta}{[(\bar{t}'/\beta)^2 - \bar{x}^2]^{1/2}}\right\}$$

(B-2)

for $\bar{t}' \geq 0$, which is Eq. (3b). Equation (3a) is derived in Eq. (A-3).

## Appendix C: Phase-locking Condition in the Relativistic Regime

As in the nonrelativistic case, we ignore image charge effects. In addition, we assume that the electric field is $-\tilde{V}_{RF}/d$ for simplicity. The relativistic force law reads,

$$\frac{dp}{dt} = m\frac{d(\gamma \dot{z})}{dt} = \frac{e\tilde{V}_{RF}}{d}\sin(\omega t + \theta)$$

(C-1)

where $\gamma = (1 - \dot{z}^2/c^2)^{-1/2}$. Integrating both sides, assuming $\dot{z}(t = 0) = 0$, we obtain

$$\gamma \dot{z} = \zeta(t) \equiv -\frac{e\tilde{V}_{RF}}{m\omega d}(\cos(\omega t + \theta) - \cos\theta)$$

(C-2)

which may be rearranged to read

$$\dot{z} = \frac{dz}{dt} = \frac{\zeta(t)}{(1+\zeta^2(t)/c^2)^{1/2}}.$$

(C-3)

Integrating again, we obtain

$$z - d = \int_0^t dt \frac{\zeta(t)}{(1+\zeta^2(t)/c^2)^{1/2}}.$$

(C-4)

Since $z = 0$ when $\omega t = \pi$, Eqs. (C-2) and (C-4) become

$$\zeta(t) = -\frac{e\tilde{V}_{RF}}{m\omega d}(-\cos(\omega t) + 1)$$

(C-5)

$$\int_0^{\pi/\omega} dt \frac{-\zeta(t)}{(1+\zeta^2(t)/c^2)^{1/2}} = d$$

(C-6)

where we have set the launch angle $\theta = \pi$. Equations (C-5) and (C-6) can be used to determine $\omega$ as a function of $\tilde{V}_{RF}$, which is shown in the solid blue curve in Fig. 5. From Eqs. (C-3) and (C-5), we may write the impact speed as

$$\beta = \frac{\frac{2e\tilde{V}_{RF}}{mc\omega d}}{\left(1 + \left(\frac{2e\tilde{V}_{RF}}{mc\omega d}\right)^2\right)^{1/2}}.$$

(C-7)

Clearly, in the nonrelativistic limit, $c \to \infty$, $\beta = 2e\tilde{V}_{RF}/mc\omega d$. This nonrelativistic limit, and Eq. (C-7), are shown in the orange curves in Fig. 5.